\documentclass{article}
\usepackage{dcase2021_techrep,amsmath,graphicx,url,times,booktabs, tabularx, subcaption, caption}
\usepackage{adjustbox}
\usepackage{amssymb}
\usepackage{kotex}

\usepackage{xcolor}         
\usepackage{multirow}
\newcommand\sh[1]{\textcolor{black} {#1}} 
\newcommand\sm[1]{\textcolor{black} {#1}} 
\title{QTI Submission to DCASE 2021: residual normalization for device-imbalanced acoustic scene classification with efficient design}

\name{Byeonggeun Kim$^1$, Seunghan Yang$^1$, Jangho Kim$^{1,2,*}$, Simyung Chang$^1$}
\address{$^1$Qualcomm AI Research${}^{\dagger}$\thanks{  ${}^{\dagger}$ Qualcomm AI Research is an initiative of Qualcomm Technologies, Inc.${}^{*}$Author completed the research in part during an internship at Qualcomm Technologies, Inc.}, Qualcomm Korea YH, Seoul, Republic of Korea\\
$^2$Seoul National University, Seoul, Republic of Korea\\
      \{kbungkun, seunghan, jangkim, simychan\}@qti.qualcomm.com} 

\begin{document}

\ninept
\maketitle

\begin{sloppy}

\begin{abstract}
This technical report describes the details of our TASK1A submission of the DCASE2021 challenge. The goal of the task is to design an audio scene classification system for device-imbalanced datasets under the constraints of model complexity. This report introduces four methods to achieve the goal. First, we propose Residual Normalization, a novel feature normalization method that uses instance normalization with a shortcut path to discard unnecessary device-specific information without losing useful information for classification. Second, we design an efficient architecture, BC-ResNet-Mod, a modified version of the baseline architecture with a limited receptive field. Third, we exploit spectrogram-to-spectrogram translation from one to multiple devices to augment training data. Finally, we utilize three model compression schemes: pruning, quantization, and knowledge distillation to reduce model complexity. The proposed system achieves an average test accuracy of 76.3\% in TAU Urban Acoustic Scenes 2020 Mobile, development dataset with 315k parameters, and average test accuracy of 75.3\% after compression to 61.0KB of non-zero parameters. We extend this work to \cite{RFN}.
\end{abstract}


\section{Introduction}
\label{sec:intro}
Acoustic scene classification (ASC) is the task of classifying sound scenes such as ``airport'', ``train station'', and ``urban park'' to which a user belongs. ASC is an important research field that plays a key role in various applications such as context-awareness and surveillance \cite{valenti2016dcase,radhakrishnan2005audio,chu2009environmental}. Detection and Classification of Acoustic Scenes and Events (DCASE) \cite{dcase2021web} is an annual challenge, attracting attention to the field and encouraging research. There are various interesting tasks in the DCASE2021 challenge, and we aim for \sh{TASK1A:} Low-Complexity Acoustic Scene Classification with Multiple Devices \cite{dcase_task1A, dcase_dataset}.

TASK1A classifies \sm{ten} different audio scenes from 12 European cities using \sm{four} different devices and 11 additional simulated devices. This year, the task becomes more challenging as an ASC model needs to meet two goals simultaneously; First, data is collected from multiple devices\sm{,} and the number of samples per device is unbalanced. Therefore, the proposed system needs to solve the domain imbalance problem while \sh{generalizing to different devices.} 
Second, TASK1A restricts the model size and therefore requires an efficient network design.

In recent years, a number of researches have been proposed for more efficient and high-performance ASC.
\sh{Most of them are based on convolutional neural network (CNN)}
using residual network and ensemble \cite{task1a2020best_cnn, receptivefield, task1a2020_2nd_cnn, task1a2019best}.
The top-performing models in the previous TASK1A utilize multiple CNNs in a single model with parallel connections \cite{task1a2020best_cnn, task1a2020_2nd_cnn}.
For the generalization of the model, \cite{receptivefield, phaye2019subspectralnet} show that there is a regularization effect by adjusting the receptive field size in CNN-based design.
However, these works also use models of several MB, and it is still challenging to satisfy the low model complexity of TASK1A of this year. In addition, when using the previous methods, we found an accuracy drop of up to 20\% on the unseen devices compared to the device with sufficient training data. In this work, we propose methods to leverage the generalization capabilities of unseen devices while maintaining the model's performance in lightweight models. First, we introduce a network architecture for ASC that utilizes broadcasted residual learning \cite{bcresnet}. Based on this architecture, we can achieve higher accuracy while reducing the size by a third of the baseline \cite{receptivefield}. Next, we propose a novel normalization method, Residual Normalization (ResNorm), \sh{which} can leverage the generalization performance for unseen devices. ResNorm allows maintaining classification accuracy while minimizing the influence on different frequency responses of devices by performing normalization of frequency bands in the residual path. 
Furthermore, we introduce a device translator that converts data from one to other devices. With the aid of data augmentation with the translator, we can mitigate the domain imbalance problem more as well as the domain difference among multiple devices. Finally, we describe model compression combined with pruning and quantization to satisfy the model complexity of the task while maintaining performance using knowledge distillation.

The rest of the report is organized as follows. Section 2 describes the network architecture, Residual Normalization, device translator, and model compression methods. Section 3 shows the experimental results and analysis. Finally, we conclude the work in Section 4.

\section{Proposed Method}
\label{sec:method}
This session introduces an efficient model design for the task of device-imbalanced acoustic scene classification. First, we \sm{present} a modified version of Broadcasting residual network \cite{bcresnet} for the acoustic scene domain. Following, we propose Residual Normalization for generalization in a device-imbalanced dataset. Next, we explain a spectrogram-to-spectrogram translation network to convert data from one device to other devices. Finally, we describe how to get a compressed version of the proposed system.

\subsection{Network Architecture}
\label{sec:architecture}

To design a low-complexity network in terms of the number of parameters, we use \sm{a}  Broadcasting-residual network (BC-ResNet) \cite{bcresnet}, a baseline architecture that uses 1D and 2D CNN feature\sh{s} together for better efficiency. While the \cite{bcresnet} targets human voice, we \sh{aim to} classify the audio scenes.
\sh{Therefore, we make two modifications to the network, {\it i.e.}, limit the receptive field and use max-pool instead of dilation, to adapt to the differences in input domains.}

The proposed architecture is shown in Table~\ref{architecture}, a fully CNN named modified BC-ResNet (BC-ResNet-Mod). The model has 5x5 convolution on the front with a 2x2 stride for downsampling followed by BC-ResBlocks \cite{bcresnet}. In \cite{receptivefield}, they show that the size of the receptive field can regularize CNN-based ASC models. We change the depth of the network and use max-pool to control the size of the receptive field. With a total of 9 BC-ResBlocks and two max-pool layers, the receptive field size is 109x109. We also do the last 1x1 convolution before global average pooling that the model classifies each receptive field separately and ensembles them by averaging. BC-ResNets use dilation in temporal dimension to obtain a larger receptive field while maintaining temporal resolution across the network. We observe that time resolution does not need to be fully kept in the audio scene domain, and instead of dilation, we insert max-pool layers in the middle of the network.

\begin{table}[t]
    \caption{\textbf{BC-ResNet-Mod.} Each row is a sequence of one or more identical modules repeated $n$ times with input shape of frequency by time by channel and total time step $T$.}
    \vskip -0.05in
    \label{architecture}
    \centering
    \resizebox{\linewidth}{!}{
    \setlength{\tabcolsep}{1em}
    \begin{tabular}{c|c|c|c}
    \toprule
    Input 
    & Operator & n & Channels \\
    \midrule
    $256 \times T \times 1$ & conv2d 5x5, stride 2& - & 2c\\
    $128 \times T/2 \times 2c$ & stage1: BC-ResBlock & 2 & c\\
    $128 \times T/2 \times c$ & max-pool 2x2 & - & - \\
    $64 \times T/4 \times c$ & stage2: BC-ResBlock & 2 & 1.5c\\
    $64 \times T/4 \times 1.5c$ & max-pool 2x2 & - & - \\
    $32 \times T/8 \times 1.5c$ & stage3: BC-ResBlock & 2 & 2c\\
    $32 \times T/8 \times 2c$ & stage4: BC-ResBlock & 3 & 2.5c\\
    $32 \times T/8 \times 2.5c$ & conv2d 1x1 & - & num class\\
    $32 \times T/8 \times$ num class & avgpool & - & - \\
    $1 \times 1 \times$ num class & - & - & - \\
    \bottomrule
    \end{tabular}
    }
    \vskip -0.1in
\end{table}


\subsection{Residual Normalization}
Instance normalization (IN) \cite{instancenorm} is a representative approach to reducing unnecessary domain gaps for better domain generalization \cite{batchinstancenorm} or domain style transfer \cite{adain} in the image domain.
While domain difference can be captured by channel mean and variance in \sh{the} image domain, we observe that differences between audio devices are revealed along frequency dimension. To get audio device generalized features, we use instance normalization by frequency (FreqIN) as below.
\begin{equation}
FreqIN(x) = \frac{x-\mu_{nf}}{\sqrt{\sigma_{nf}^{2} + \epsilon}},
\label{eq:freqin}
\end{equation}
where,
\begin{align}
\mu_{nf} &= \frac{1}{CT}\sum_{c=1}^{C}\sum_{t=1}^{T}x_{ncft}, \nonumber \\ 
\sigma^2_{nf} &= \frac{1}{CT}\sum_{c=1}^{C}\sum_{t=1}^{T}{(x_{ncft}-\mu_{nf})^2}. 
\label{eq:meanvar}
\end{align}
\sm{Here, $\mu_{nf}$, $\sigma_{nf} \in \mathbb{R}^{N\times F}$ are mean and standard deviation of the input feature $x \in \mathbb{R}^{N\times C\times F\times T}$, where $N$, $C$, $F$, $T$ denote batch size, number of channel, frequency dimension, and time dimension respectively.}
\sh{$\epsilon$ is a small number added to avoid division by zero.}


Direct use of IN can result in loss of useful information for classification contained in domain information. To compensate for information loss due to FreqIN, we add an identity path multiplied by a hyperparameter $\lambda$. We suggest a normalization method, named Residual Normalization (ResNorm) which is
\begin{equation}
\textit{ResNorm}(x) = \lambda \cdot x + \textit{FreqIN}(x).
\label{eq:resnorm}   
\end{equation}
We apply ResNorm for input features and after the end of every stage in Table~\ref{architecture}. There are a total of five ResNorm modules in the network.

\begin{table*}[t]
    \caption{\textbf{Overall test results.} Top-1 test accuracy (\%) and standard deviation on TAU Urban AcousticScenes 2020 Mobile, development dataset for each device (A to S6). Device S4, S5 and S6 remain unseen in training (averaged over 3 seeds, * averaged over 6 seeds).}
    \vskip -0.05in
    \label{result_table}
    \centering
    \resizebox{\linewidth}{!}{
    \begin{tabular}{l|c|ccccccccc|c}
    \toprule
    Method & \#Param & A & B & C & S1 & S2 & S3 & S4 & S5 & S6 & Overall  \\
    \midrule
    BC-ResNet-Mod-1 & 8.1k & 73.1 & 61.2 & 65.3 & 58.2 & 57.3 & 66.2 & 51.5 & 51.5 & 46.3 & 58.9 $\pm$ 0.8 \\
    BC-ResNet-Mod-1 + Global FreqNorm & 8.1k & 73.9 & 60.9 & 65.5 & 60.2 & 57.9 & 67.9 & 50.2 & 54.3 & 49.4 & 60.0 $\pm$ 0.9\\
    BC-ResNet-Mod-1 + FreqIN & 8.1k & 69.9 & 63.5 & 60.0 & 65.3 & \textbf{66.7} & 67.6 & \textbf{65.9} & 64.9 & 62.0 & 65.1 $\pm$ 0.6\\
    BC-ResNet-Mod-1 + Pre-ResNorm & 8.1k & 75.1 & \textbf{68.9} & 67.0 & \textbf{66.0} & 63.9 & 69.3 & 63.4 & \textbf{66.9} & \textbf{63.6} & \textbf{67.1 $\pm$ 0.8}\\
    BC-ResNet-Mod-1 + ResNorm & 8.1k & \textbf{76.4} & 65.1 & \textbf{68.3} & \textbf{66.0} & 62.2 & \textbf{69.7} & 63.0 & 63.0 & 58.3 & *65.8 $\pm$ 0.7\\
    \midrule
    CP-ResNet, c=64 & 899k & 77.0 & 69.3 & 69.6 & 70.3 & 68.2 & 70.9 & 62.7 & 63.9 & 58.1 & 67.8\\
    BC-ResNet-8, $\text{num SSN group}=4$ & 317k & 77.9 & 70.4 & 72.4 & 69.5 & 68.3 & 69.8 & 66.3 & 64.1 & 58.6 & 68.6 $\pm$ 0.4\\
    BC-ResNet-Mod-8 & 315k & 80.7 & 72.8 & \textbf{74.4} & 71.4 & 68.7 & 71.0 & 62.2 & 65.3 & 59.4 & 69.5 $\pm$ 0.3\\
    BC-ResNet-Mod-8 + Pre-ResNorm & 315k & 80.8 & 73.7 & 73.0 & 74.0 & 72.9 & 77.8 & \textbf{73.3} & 72.1 & 71.0 & 74.3 $\pm$ 0.3 \\
    BC-ResNet-Mod-8 + ResNorm & 315k & \textbf{81.3} & \textbf{74.4} & 74.2 & \textbf{75.6} & \textbf{73.1} & \textbf{78.6} & 73.0 & \textbf{74.0} & \textbf{72.7} & \textbf{*75.2 $\pm$ 0.4} \\
    \midrule
    BC-ResNet-Mod-8 + ResNorm, Device Translator & 315k & 80.5 & 74.4 & 73.9 & 76.0 & 73.2 & 78.5 & 74.1 & 74.1 & 73.6 & 75.4 $\pm$ 0.3 \\
    BC-ResNet-Mod-8 + ResNorm, 300epoch, KD & 315k & 82.6 & 75.6 & 74.7 & 77.0 & 74.2 & 78.7 & 75.1 & 74.8 & 73.4 & *76.3 $\pm$ 0.8 \\
    ~~ + model compress & - & 82.0 & 73.8 & 74.3 & 76.2 & 73.2 & 78.8 & 73.8 & 72.8 & 73.3 & *75.3 $\pm$ 0.8 \\
    \bottomrule
    \end{tabular}
    }
    \vskip -0.1in
\end{table*}

\begin{figure}
    \centering
    \begin{subfigure}[t]{0.22\textwidth}
        \centering
        \includegraphics[width=1.0\textwidth]{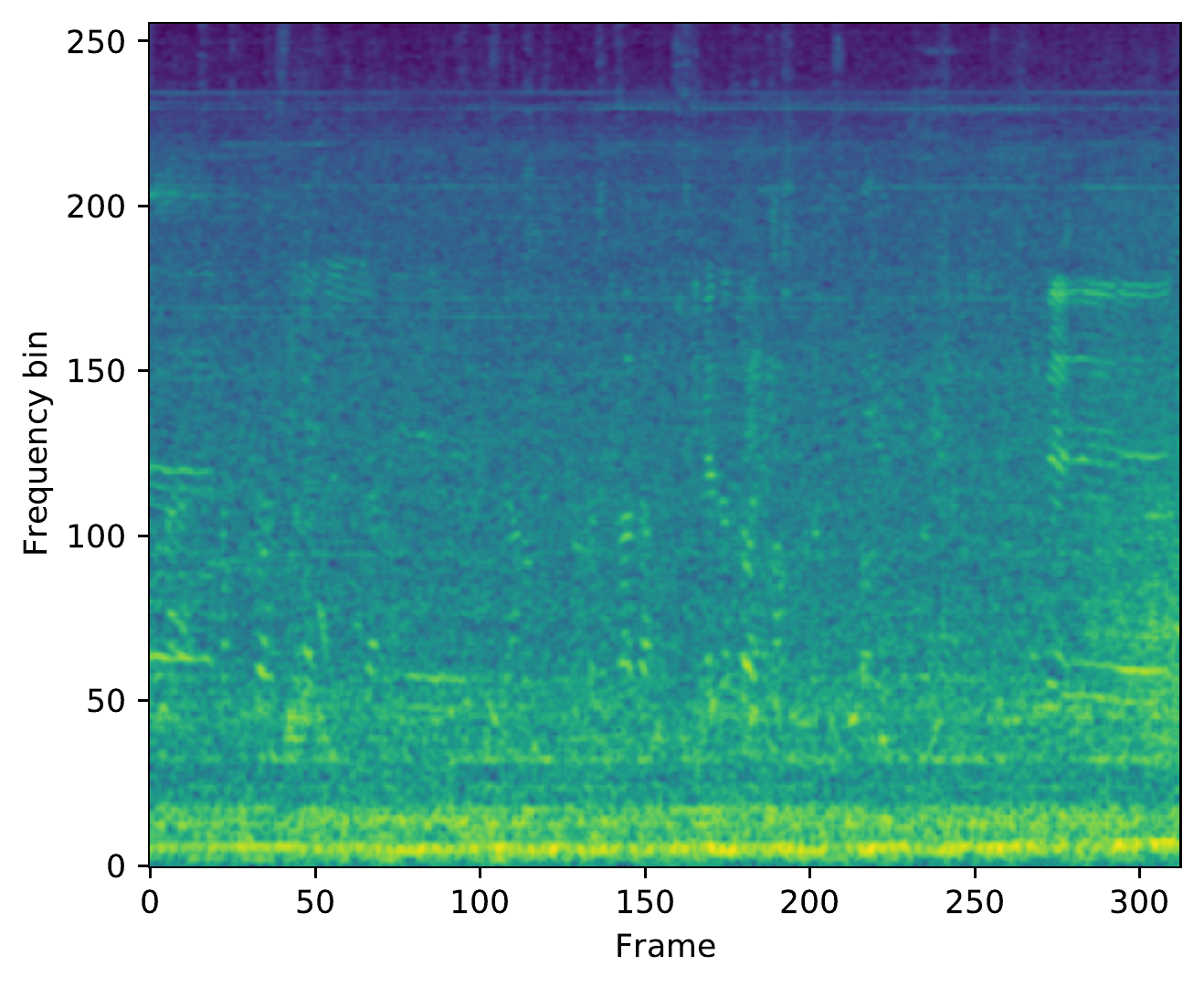}
        \vskip -0.05in
        \caption{{\small Real ``a''}}    
    \end{subfigure}
    \hfill
    \begin{subfigure}[t]{0.22\textwidth}  
        \centering 
        \includegraphics[width=1.0\textwidth]{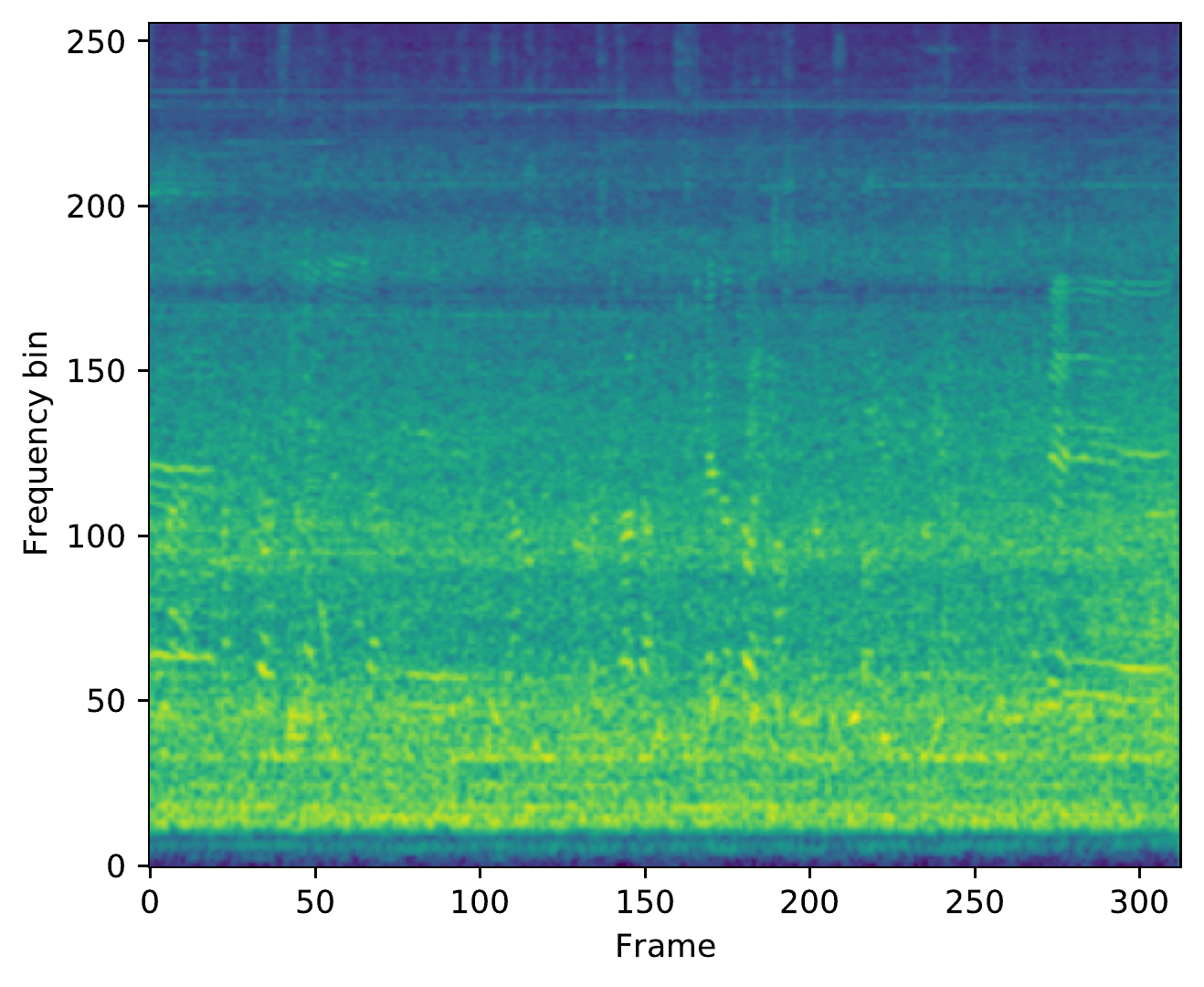}
        \vskip -0.05in
        \caption{{\small Translated ``s1''}}    
    \end{subfigure}
    \begin{subfigure}[t]{0.22\textwidth}   
        \centering 
        \includegraphics[width=1.0\textwidth]{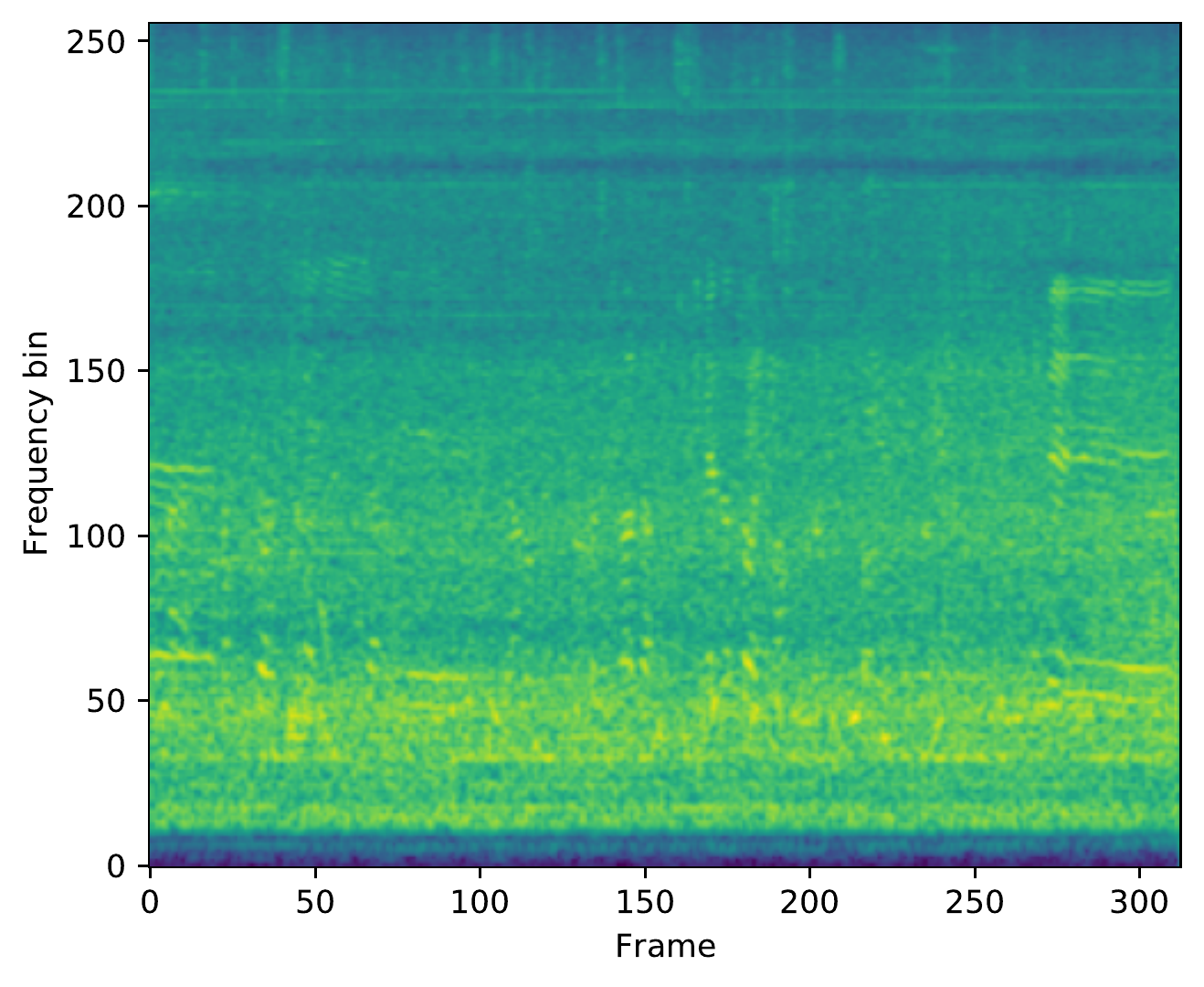}
        \vskip -0.05in
        \caption{{\small Translated ``s2''}}    
    \end{subfigure}
    \hfill
    \begin{subfigure}[t]{0.22\textwidth}   
        \centering 
        \includegraphics[width=1.0\textwidth]{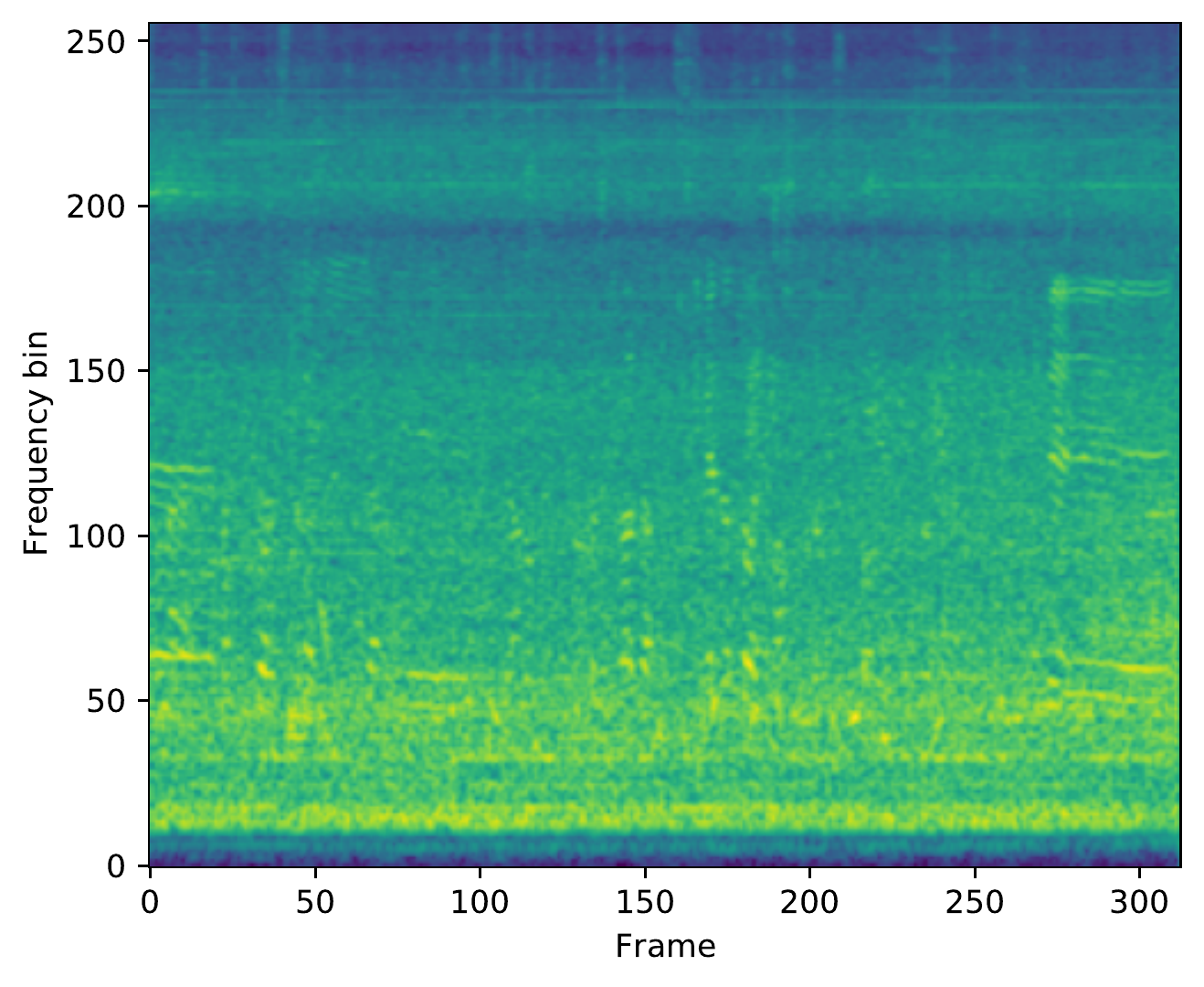}
        \vskip -0.05in
        \caption{{\small Translated ``s3''}}    
    \end{subfigure}
    \caption{\textbf{Examples of device translation.} (b), (c), and (d) are generated by the device translator using the real data (a).} 
    \label{translation_example}
    \vspace{-3mm}
\end{figure}

\subsection{Device Translator}
We design the device translator, which converts a spectrogram from one device into another. We start with the pix2pix framework~\cite{isola2017image}, which has been shown to produce high-quality results in image-to-image translation.
As in pix2pix, we exploit an encoder-decoder network and a discriminator, but we redesign the encoder-decoder suitable for the audio domain.
To consider the characteristics of audio, we adapt Subspectral Normalization~\cite{ssn} to the encoder-decoder network.
In addition, we observe that the U-Net with a large depth can reconstruct the spectrogram well.
Therefore, we use the U-Net architecture~\cite{ronneberger2015u} as the encoder-decoder and insert several residual blocks in the middle of the network.

We train the network with two objectives, a reconstruction loss and a GAN-based loss.
We use L2 distance as a reconstruction loss for synthesizing the spectrogram. It helps the network to produce similar output to the known paired ground truth image.
In addition, we utilize LSGAN~\cite{mao2017least} as a GAN-based loss that is designed to encourage the network generating the image to be realistic and diverse.
With these two loss functions, the device translator can convert the spectrogram from one to other devices.

\subsection{Model Compression}
\label{model_compression}
To compress the proposed model, we utilize three model compression schemes: pruning, quantization, and knowledge distillation.

\noindent \textbf{Pruning.} The pruning method prunes unimportant weights or channels based on many criteria. In this work, we choose a magnitude-based one-shot unstructured pruning scheme used in \cite{NEURIPS2020_eb1e7832}. \sm{After training, we conduct unstructured pruning on all convolution layers and do additional training to enhance the pruned model's performance.}

\noindent \textbf{Quantization.} 
\sm{Quantization is the method to map continuous infinite values to a smaller set of discrete finite values.}
We quantize all of our models with quantization-aware training (QAT) with symmetric quantization \cite{NEURIPS2020_eb1e7832}. We combine the pruning and quantization methods. So, we quantize the important weights which are not pruned after the pruning process in the additional training phase. Basically, we utilize the half-precision representation for all weights. For some models, we quantize all convolution layers as an 8-bit, shown in Table \ref{table:model_var}.

\noindent \textbf{Knowledge Distillation.} Knowledge Distillation (KD) trains the lightweight model using a pre-trained teacher network. In general, previous model compression schemes such as pruning and quantization decrease the performance by reducing the model complexity. To enhance the performance of the compressed model, we use a KD loss \cite{kim2021feature} using the pre-trained model as a teacher network. 

\section{Experiments}
\label{sec:experiment}
\subsection{Experimental Setup}
\noindent \textbf{Datasets.} We evaluate the proposed method on the TAU Urban Acoustic Scenes 2020 Mobile, development dataset \cite{dcase_dataset}. The dataset consists of a total of 23,040 audio segment recordings from 12 European cities in 10 different acoustic scenes using 3 real devices (A, B, and C) and 6 simulated devices (S1-S6). The 10 acoustic scenes contain ``airport'', ``shopping mall'', ``metro station'', ``pedestrian street'', ``public square'', ``street with traffic'', ``park'', and travelling by ``tram'', ``bus'', and ``metro''. Audio segments from B and C are recorded simultaneously with device A, but not perfectly synchronized.
Simulated devices S1-S6 generate data using randomly selected audio segments from real device A.
Each utterance is 10-sec-long and the sampling rate is 48kHz.
\cite{dcase_dataset} divides the dataset into training and test of 13,962 and 2,970 segments, respectively. In the training data, device A has 10,215 samples while B, C, and S1-S3 have 750 samples each, which means the data is device\sh{-}imbalanced. Devices S4-S6 remain unseen in training. In test data, all devices from A to S6 have 330 segments each.
For the device translator, we use paired data between devices existing in the dataset. 

\noindent \textbf{Implementation Details.} We do downsampling by 16kHz and use input features of 256-dimensional log Mel spectrograms with a window length of 130ms and a frameshift of 30ms. During training, we augment data to get a more generalized model. In the time dimension, we randomly roll each input feature in the range of -1.5 to 1.5 sec, and the out-of-range part is added to the opposite side. We also use Mixup \cite{mixup} with $\alpha=0.3$ and Specaugment \cite{specaugment} with two frequency masks and two temporal masks with mask parameters of 40 and 80\sm{, respectively, }except time warping. In BC-ResNet-Mod, we use Subspectral Normalization \cite{ssn} as indicated in \cite{bcresnet} with 4 sub-bands and use dropout rate of 0.1. We train the models for 100 epoch using stochastic gradient descent (SGD) optimizer with momentum to 0.9, weight decay to 0.001, mini-batch size to 64, and learning rate linearly increasing from 0 to 0.06 over the first five epochs as \sm{a} warmup \cite{warmup} before decaying to zero with cosine annealing \cite{cosine_schedule} for the rest of the training.
For the device translator, we use the U-Net architecture and insert six residual blocks in the middle of the network with Subspectral Normalization with 16 sub-bands.
We follow the same training configurations used in~\cite{isola2017image}.

\noindent \textbf{Baselines.} We compare our method with simple baselines and some prior works: 1) Global FreqNorm, which normalizes data by global mean and variance of each frequency bin;
2) Pre-ResNorm, which uses ResNorm as preprocessing module; 3) FreqIN, which is a special case of Pre-ResNorm when $\lambda=0$ in equation~\ref{eq:resnorm}; 4) BC-ResNet-8, num SSN group=4, which is original BC-ResNet-8 in \cite{bcresnet} with the number of subspectral normalization group of 4 \cite{ssn};  5) CP-ResNet \cite{receptivefield}, which is a residual network-based ASC model with limited receptive field size.

\begin{table*}[t]
    \caption{\textbf{Submitted Models.} We get four variants by considering use of entire development set, device translator, model selection method, pruning ratio, bits per parameter, and number of ensemble. The symbols `O' and `X' imply that we utilize certain method or not, respectively. \sm{In addition, `Official Split' denotes the test performance when we train each system with provided the cross-validation setup.}}
    \label{table:model_var}
    \centering
    \resizebox{\linewidth}{!}{
    \begin{tabular}{c|cc|cccccc|cc|cc}
    \toprule
    
    \multirow{2}{*}{Model} & \multirow{2}{*}{Full Dev. data} & \multirow{2}{*}{Translator} & Model & Prune Ratio & Bits per & \multirow{2}{*}{Ensemble} & \#Params & \multirow{2}{*}{Size} & \multicolumn{2}{c|}{Submitted} &  \multicolumn{2}{c}{Official Split} \\
    & & & Select & (Conv) & Param & & (total/non-zero) & & Acc (\%) & LogLoss & Acc (\%) & LogLoss \\
    \midrule
    
    \multirow{2}{*}{1} & teacher (O) & \multirow{2}{*}{X} & \multirow{2}{*}{Min Var} & \multirow{2}{*}{89\%} & 8-bit (conv) & \multirow{2}{*}{2} & \multirow{2}{*}{630k/95k} & \multirow{2}{*}{121.9KB} & \multirow{2}{*}{81.06} & \multirow{2}{*}{0.56}&
    \multirow{2}{*}{76.99} & \multirow{2}{*}{0.72} \\
    & student (X) & & & & + 16-bit & & & & &\\
    
    \midrule
    
    \multirow{2}{*}{2} & teacher (w/o test set) & \multirow{2}{*}{O} & \multirow{2}{*}{Last} & \multirow{2}{*}{89\%} & 8-bit (conv) & \multirow{2}{*}{2} & \multirow{2}{*}{630k/95k} & \multirow{2}{*}{121.9KB} & \multirow{2}{*}{86.28} & \multirow{2}{*}{0.40}&
    \multirow{2}{*}{75.91} & \multirow{2}{*}{0.72} \\
    & student (w/o test set) & & & & + 16-bit & & & & &\\
    
    \midrule
    
    \multirow{2}{*}{3} & teacher (O) & \multirow{2}{*}{X} & \multirow{2}{*}{Last} & \multirow{2}{*}{89\%} & 8-bit (conv) & \multirow{2}{*}{2} & \multirow{2}{*}{630k/95k} & \multirow{2}{*}{121.9KB} & \multirow{2}{*}{-} & \multirow{2}{*}{-} &
    \multirow{2}{*}{77.46} & \multirow{2}{*}{0.72}\\
    & student (O) & & & & + 16-bit & & & & &\\
    
    \midrule
    
    \multirow{2}{*}{4} & teacher (O) & \multirow{2}{*}{O} & \multirow{2}{*}{Last} & \multirow{2}{*}{84\%} & \multirow{2}{*}{16-bit} & \multirow{2}{*}{X} & \multirow{2}{*}{313k/63k} & \multirow{2}{*}{122.5KB} & \multirow{2}{*}{82.17} & \multirow{2}{*}{0.51} & \multirow{2}{*}{75.10} & \multirow{2}{*}{0.77} \\
    & student (X) & & & & & & & & &\\
    
    \bottomrule
    \end{tabular}
    }
    \vskip -0.1in
\end{table*}

\subsection{Results and Discussion}

Table~\ref{result_table} compares baselines with our \textit{BC-ResNet-Mod-1}  and \textit{BC-ResNet-Mod-8} whose base number of channels $c$ are 10 and 80 respectively in Table~\ref{architecture}. We use fixed $\lambda = 0.1$ for Pre-ResNorm and ResNorm in experiments.

\noindent \textbf{Network architecture.} We compare our BC-ResNet-Mod-8 with BC-ResNet-8 \cite{bcresnet} and CP-ResNet \cite{receptivefield}. As shown in Table~\ref{result_table}, BC-ResNet-Mod-8 records Top-1 test accuracy 69.5\% with only one-third number of parameters compared to CP-ResNet showing 67.8\% accuracy.
Moreover, BC-ResNet-Mod-8 outperforms the original BC-ResNet-8 by a 1\% margin with modifications.

\noindent \textbf{Impact of Residual Normalization.} Residual normalization plays an important role in generalization to device imbalanced dataset. First, we compare ResNorm with the baselines in BC-ResNet-Mod-1. Baselines, Global FreqNorm and FreqIN are preprocessing modules. While Global FreqNorm poorly generalizes \sh{to} unseen devices S4-S6, 
FreqIN-based models generalize better and achieve a 5\% margin compared to the Global FreqNorm. However, FreqIN shows degradation in seen devices A-S3 due to strong regularization.
\sh{If we apply ResNorm as a preprocessing module (Pre-ResNorm) with a fixed $\lambda=0.1$, the model can achieve 67.1\% test accuracy.}
Pre-ResNorm shows higher test accuracy for seen domains A-S3 \sh{than FreqIN} while maintaining good generalization for unseen devices. The results imply that ResNorm discards the unnecessary domain information by IN while keeping the useful information through the connection, $\lambda \cdot x$ in equation~\ref{eq:resnorm}.
The BC-ResNet-Mod-8 is improved by a large margin using ResNorm and achieves average test accuracy of 75.2\%. Due to its regularization effect, ResNorm depresses the performance of the small model, BC-ResNet-Mod-1. In Table~\ref{result_table}, Pre-ResNorm which uses ResNorm once at preprocessing works better than ResNorm for the small model while ResNorm works better for the large model, BC-ResNet-Mod-8.

\noindent \textbf{Impact of spectrogram-to-spectrogram translation.}
In Fig. \ref{translation_example}, we illustrate spectrograms translated by the real data of device a.
Each spectrogram contains the characteristic of its device.
We convert all training data and train our scene classification model with both original and translated datasets.
As shown in Table~\ref{result_table}, the performance variation across the devices decreases, and it indicates that training with translated data can improve the domain generalization ability of the network.



\subsection{Submitted Models}
\sh{To get higher accuracy, we train the network using the best method, BC-ResNet-Mod-8 with ResNorm\sm{,} for a total of 300 epochs with 10 epochs of warmup.
It achieves average test accuracy of 75.3\%. Also, we utilize the trained model as the teacher and train the newly initialized model by performing knowledge distillation. \sm{Then our model} achieves 76.3\%.}
We compress the model by 89\% pruning, 8-bit quantization, and half-precision for convolution layers and other layers, respectively, and it achieves 75.3\% as in Table~\ref{result_table}.

Task1A allows submitting four different models.
\sh{We choose various models as shown in Table~\ref{table:model_var}.}
\sh{Our objective is to train the models performing well on the TAU Urban AcousticScenes 2021 Mobile, evaluation dataset \cite{dcase_dataset} whose labels are unknown.}

\sh{For models ``1" and ``4", we utilize the full development set for the teacher models, but the students are trained on the original cross-validation setup. We train the teacher and student on the full development set without the cross-validation test set for model ``2". Only model ``3" uses full development set for both teacher and student models.}
We use two methods for model selection: ``Last" is the model at the end of training, and ``Min Var" means the model having minimum performance differences among devices.
In addition, we adopt the following variations: 1) doing 8-bit quantization and ensemble two models with different seeds or using a single half-precision model and 2) using device translator or not. \textit{Model 1, 2,} and \textit{3} have 66.1k 8-bit nonzero for convolution layers and 29.4k 16-bit parameters for normalization, resulting in 121.9KB. \textit{Model~4} has 48.0k nonzero parameters for convolution layers and 14.7k for others. And the model size is 122.5KB in total with 16-bit half-precision.

\section{Conclusions}
\label{sec:conclude}

In this work, we design a system to achieve two goals; 1) efficient design in terms of the number of parameters and 2) adapting to device imbalanced dataset. To design an efficient acoustic scene classification model, we suggest a modified version of Broadcasting residual network \cite{bcresnet} by limiting receptive field and using max-pool. We compress the model further by utilizing three model compression schemes, pruning, quantization, and knowledge distillation. Also, we suggest a frequency-wise normalization method, named Residual Normalization which uses instance normalization by frequency and shortcut connection to be generalized to \sm{multiple devices} while not losing \sm{discriminative} information. Our system achieves 76.3\%
test accuracy on TAU Urban \sm{Acoustic Scenes 2020 Mobile}, development dataset with 315k number of parameters and the compressed version \sm{achieves} 75.3 \% test accuracy with 89\% pruning, 8-bit quantization, and knowledge distillation.
We extend this work to \cite{RFN}.

\bibliographystyle{IEEEtran}
\bibliography{main}

\end{sloppy}
\end{document}